\documentclass[aps,prl,reprint,a4paper,longbibliography,footinbib,superscriptaddress]{revtex4-1}
\usepackage[latin9]{inputenc}
\usepackage{graphicx}
\usepackage{siunitx}
\usepackage{xcolor}
\usepackage[normalem]{ulem}
\usepackage{bm}

\newcommand  {\at}   {\mathrm{at}}
\newcommand  {\sub}  {\mathrm{sub}}

\newcommand  {\D}  {\mathrm{D}}
\newcommand  {\kB}  {k_\mathrm{B}}
\newcommand  {\mRb}  {m_\mathrm{Rb}}

\newcommand {\INPHYNI} {Universit\'e C\^ote d'Azur, CNRS, Institut de Physique de Nice, France}
\newcommand {\CAPES} {CAPES Foundation, Ministry of Education of Brazil, Bras\'{i}lia -- DF 70040-020, Brazil}

\begin{document}

\title{Robustness of Dicke subradiance against thermal decoherence}

\author{P. Weiss}
\email{patrizia.weiss@inphyni.cnrs.fr}
\affiliation{\INPHYNI}
\author{A. Cipris}
\affiliation{\INPHYNI}
\author{M. O. Ara\'{u}jo}
\email{Present address: Departamento de F\'isica, Universidade Federal de Pernambuco, 50670-901, Recife -- PE, Brazil}
\affiliation{\INPHYNI}
\affiliation{\CAPES}
\author{R. Kaiser}
\affiliation{\INPHYNI}
\author{W. Guerin}
\affiliation{\INPHYNI}

\begin{abstract}
Subradiance is the cooperative inhibition of the radiation by several emitters coupled to the same electromagnetic modes. It was predicted by Dicke in 1954 and only recently observed in cold atomic vapors. Here we address the question to what extent this cooperative effect survives outside the limit of frozen two-level systems by studying the subradiant decay in an ensemble of cold atoms as a function of the temperature. Experimentally, we observe only a slight decrease of the subradiant decay time when increasing the temperature up to several millikelvins, and in particular we measure subradiant decay rates that are much smaller than the Doppler broadening. This demonstrates that subradiance is surprisingly robust against thermal decoherence. The numerical simulations are in good agreement and allow us to extrapolate the behavior of subradiance at higher temperatures.
\end{abstract}

\maketitle

\section{Introduction}
Understanding the influence of decoherence or dephasing processes in cooperative effects such as super- and subradiance\,\cite{Dicke1954,Scully2009a,Goban2015,Guerin2016a,Araujo2016,Roof2016,Solano2017} is not only interesting from a fundamental point of view, but it is also important for the possible developments of photonic device exploiting cooperativity in the classical or quantum regime~\cite{Ostermann2013,Scully2015,Facchinetti2016,Asenjo2017,Facchinetti2018,Guimond2019}. This is especially true if one wants to use solid-state devices~\cite{Mlynek2014,Jenkins2017}, which are subject to phonon-induced decoherence.
Previous theoretical studies of various toy models in the framework of open quantum systems have predicted some robustness of superradiance to noise and dephasing\,\cite{Celardo2014a,Celardo2014b,Tayebi2016,Damanet2016}. 
 
In this article, we report an experimental study of thermal decoherence of subradiant Dicke states in a large ensemble of cold atoms\,\cite{Guerin2016a}. Indeed, even if laser-cooled atoms are not coupled to phonons, they are not completely frozen. Atomic motion has been shown to be a source of decoherence for coherent back scattering~\cite{Labeyrie2006} and, to suppress the effect of recurrent scattering on the refractive index of dense atomic media \cite{Javanainen2014, Jenkins2016}. Since subradiance is an interference effect involving very long time scales, it is expected to be particularly fragile.
The lifetime of subradiant states $\tau_\text{sub}$ can exceed several hundred times the single atom life-time $\tau_\text{at}$. An intuitive guess how residual atomic motion due to finite temperature should limit this lifetime is given by the restriction that the atoms should not move during this life time farther than the wavelength $\lambda/2\pi$ to not change the interference condition. 
Taking the rms velocity of a Maxwell-Boltzmann distribution $\sigma_v$, subradiance would require $\Gamma_\D\leq\Gamma_\sub$, where $\Gamma_\D=k\sigma_v$ is the Doppler width.

Contrary to that, and quite surprisingly, we show here that subradiance in the linear-optics regime is robust against thermal motion. We observe only a slight decrease of the subradiant decay time when increasing the temperature up to several millikelvins, and in particular we measure subradiant decay rates $\Gamma_\sub$ that are much smaller than the Doppler broadening $\Gamma_\D$. We also perform numerical simulations showing that the breakdown of subradiance only occurs when the Doppler broadening is on the same order of magnitude as the natural lifetime of the atomic transition $\Gamma_0$. In practice, this means that subradiance can be observed and used at any ``cold-atom'' temperature and even beyond.

\begin{figure}
\centerline{\includegraphics{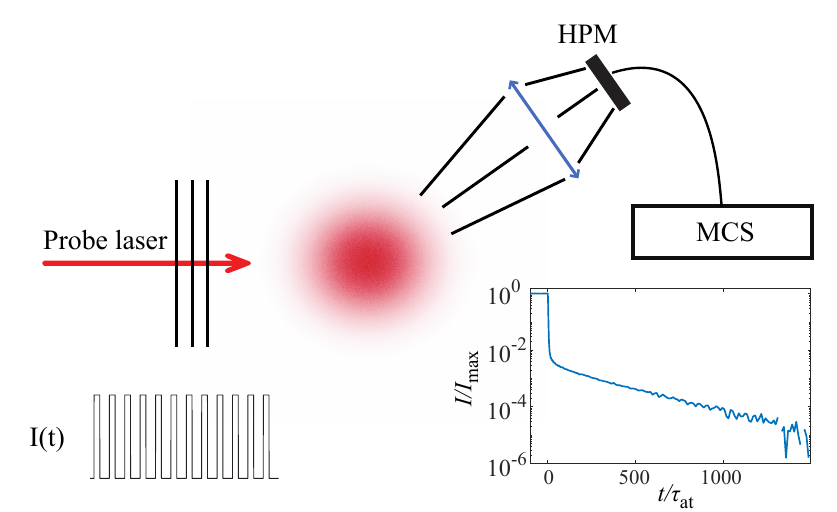}}
\caption{Sketch of our set up based on an ensemble of atoms in a MOT. The cloud is excited with a time sequence of probe laser pulses. The switch-off dynamics of the scattered light is detected under an angle \SI{35}{\degree} to the probe beam axis with a hybrid photon multiplier (HPM) and averaged with a multichannel scaler (MCS).}
\label{figsetup}
\end{figure}

\section{Experimental setup}\label{secsetup}
The experimental setup is based on a cloud of cold $^{87}$Rb atoms prepared in a magneto-optical trap (MOT). After \SI{60}{\milli \second} of loading from the background vapor and a stage of compressed MOT (\SI{30}{\milli\second}) we obtain a sample of $N\approx 3\times10^9$ atoms at a temperature $T\approx$ \SI{100}{\micro \kelvin} with a Gaussian density distribution (peak densities \mbox{$\rho_0 \sim 10^{11}$ \SI{}{\centi \meter ^{-3}}} and rms size $R \approx$ \SI{1}{\milli \meter}). A more detailed description of the setup as well as the procedure to observe and analyze subradiance can be found in \cite{Guerin2016a,Weiss2018}. For this new series of experiments we now add an optical molasses in order to vary the temperature in a controlled manner. To do so, we varied the detuning of the cooling laser in a range between $-10\Gamma_0$ to the atomic resonance. We also use the molasses duration (\SIrange{1}{10}{\milli \second}) as a parameter to tune the final temperature, which is between \SI{50}{\micro \kelvin} and \SI{11}{\milli \kelvin}, corresponding to $\Gamma_\D/\Gamma_0$ between $0.01$ and $0.2$.

\begin{figure*}
\centerline{\includegraphics{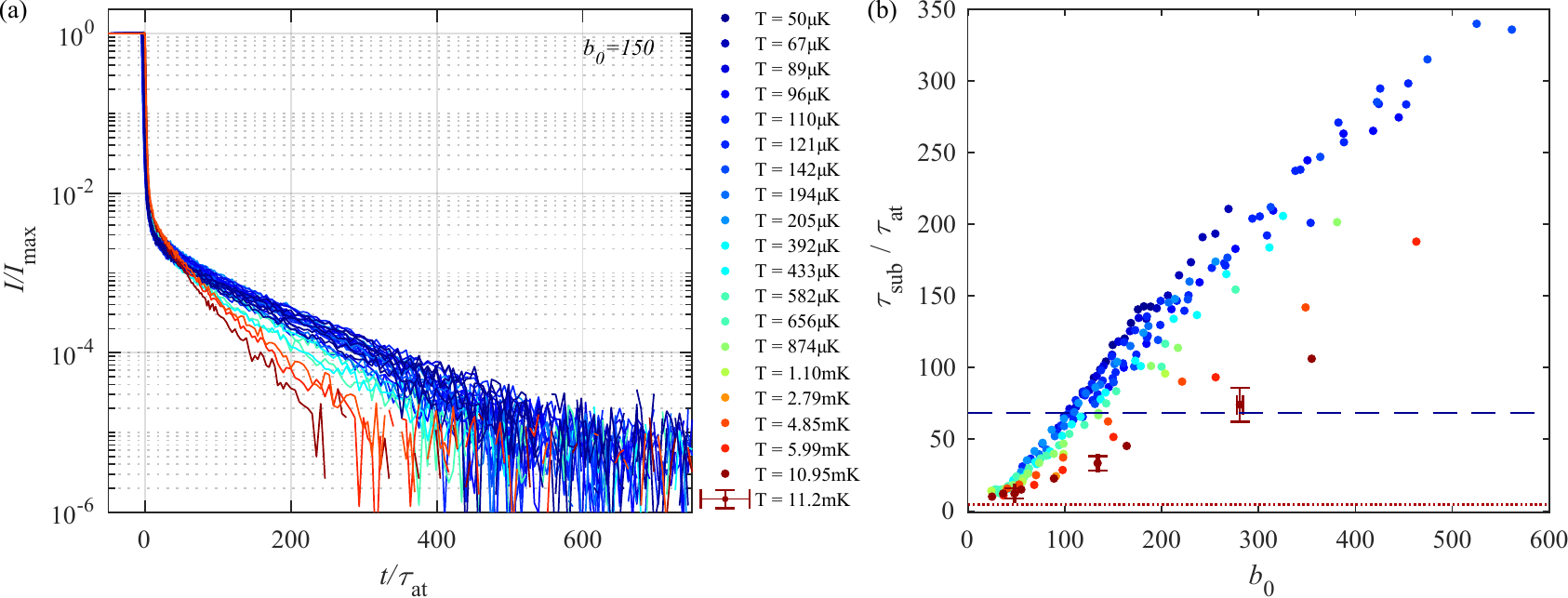}}
\caption{(a) Experimental decay curves for several temperatures, all normalized to the pulse level at the switch-off time $t=0$, with a resonant optical depth $b_0=150\pm8$. The temperature is encoded in the color scale. A smooth reduction of the subradiant decay with increasing temperature is well visible. (b) Subradiant decay times as a function of $b_0$ for different temperatures (same color code).
For clarity only the last data-set is shown with error bars. The horizontal blue dashed line shows the time scales $\Gamma_\D^{-1}$ corresponding to the Doppler width for the lowest temperature, \SI{50}{\micro \kelvin} ($\Gamma_\D^{-1}=68.6\tau_\text{at}$), and the  highest one, \SI{11.2}{\milli \kelvin} ($\Gamma_\D^{-1}=4.58\tau_\text{at}$), red dotted line.}
\label{figexp_data}
\end{figure*}
After this preparation the cloud expands ballistically. During the expansion the atoms are first optically pumped to the  hyperfine ground state $F=2$ and then excited by a series of 12 weak laser pulses with a duration of \SI{10}{\micro \second} (Fig. \ref{figsetup}). The pulses are separated by either \SI{1}{\milli \second} or \SI{0.5}{\milli \second} depending on the temperature.
The time shaping of the pulses is done with two acousto-optical modulators, achieving a faster switch-off ($t_\text{switch}\approx$ \SI{15}{\nano \second}) than the natural lifetime $\tau_\text{at} = \Gamma_0^{-1} =$ \SI{26.24}{\nano \second}, and at the same time providing an extinction ratio better than $10^{-5}$. The probe beam is large compared to the cloud size ($1/e^2$ radius \SI{5.3}{\milli \meter}) to ensure a homogeneous driving and is linearly polarized. In previous experiments it has been shown that subradiance is independent of the detuning as long as multiple scattering is negligible\,\cite{Guerin2016a,Weiss2018}. As a consequence we have chosen here to work with a constant detuning
$\delta = \left(\omega - \omega_0\right) = -4\Gamma_0$, with $\omega$ the laser frequency, $\omega_0$ the frequency of the atomic transition $F=2 \rightarrow F'=3$, and $\Gamma_0/2\pi=$ \SI{6,07}{\mega \hertz} the natural linewidth.
The intensity of the probe is chosen such that the saturation parameter is
\begin{equation}
s(\delta)=g \frac{I/I_\text{sat}}{1+4\delta^2/\Gamma_0^2}\approx 0.02 \, ,
\end{equation}
well in the linear-optics regime, with $g=7/15$ the degeneracy factor of the transition for equally populated Zeeman states and $I_\text{sat} = 1.6$ \SI{}{\milli\watt/ \centi\meter^{2}} the saturation intensity.
The scattered light is collected by a two-inch lens and detected with a hybrid photon multiplier under an angle of \SI{35}{\degree} to the probe beam (see Fig.\,\ref{figsetup}). The signal is then sent to a multichannel scaler for averaging over typically more than 400 000 cycles with a time resolution of \SI{1.6}{\nano \second}.

Because of the ballistic expansion during the pulse series the cloud size increases and the cooperativity parameter that controls super- and subradiance effects\,\cite{Guerin2016a,Guerin2017a}, $b_0=3N/(kR)^2$, decreases (here $k=2\pi/\lambda$ and $\lambda=$\,\SI{780}{\nano \meter} is the wavelength of the transition). For simplicity, we call this parameter the on-resonance optical depth although the actual optical depth is
\begin{equation}
b(\delta) = g \frac{b_0}{1+4\delta^2/\Gamma_0^2} \, ,
\end{equation}
which can be measured by absorption imaging. This measurement is interlaced with the data acquisition by changing one out of 750 cycles. For those special cycles, the measurement sequence with the pulse series is replaced by an absorption imaging procedure. The time of flight before imaging is varied over the data acquisition such that the temperature and the optical depth for each pulse are measured several times during an acquisition run. This improved calibration procedure gives us access to any drift that might occur during the data acquisition.

\section{Subradiant decay for different ensemble temperatures} \label{secsub}
We report in Fig.\ref{figexp_data} the results of our systematic experimental study of the subradiant decay as a function of the temperature of the sample. In panel (a) we show the decay curves of the scattered light. The intensity is normalized to its steady-state value just before the switch-off at $t=0$. All curves are recorded for an on-resonant optical depth $b_0 = 150\pm 8$ and the temperature is encoded in the color scale, from \SI{50}{\micro \kelvin} (dark blue) to \SI{11.2}{\milli \kelvin} (dark red). The first observation is that subradiance is clearly visible in all curves, even in the mK range, which demonstrates its robustness against thermal motion. The second is that we do observe a reduction of the subradiant decay time when the temperature increases.

To study this effect more precisely we fit all decay curves by an exponential decay at late time (the fitting range is taken as one decade above the noise floor, to cover the longest-lived visible mode). The obtained subradiant time $\tau_\sub$ is reported in Fig.\ref{figexp_data}(b) as a function of the on-resonance optical depth $b_0$ for each temperature (same color code). Note that the number of data points is reduced for the highest temperatures because of the fast ballistic expansion of the cloud. The linear scaling of $\tau_\sub$ with $b_0$, as previously reported\,\cite{Guerin2016a,Weiss2018}, is observed at all temperatures; however, the slope is reduced when the temperature increases.

For each temperature $T$ we fit this linear trend as $\tau_\text{sub}/\tau_\at = 1 + \alpha_\sub \cdot b_0 $ to obtain the slope of the subradiant enhancement. In the inset of Fig.\,\ref{figdecay_t} we plot the slope $\alpha_\sub$ as a function of the ratio between the Doppler broadening $\Gamma_\D = k \sigma_v$, with $\sigma_v=(\kB T/\mRb)^{1/2}$ the rms width of the atomic velocity distribution ($\kB$ is the Boltzmann constant and $\mRb$ the atomic mass), and the natural decay rate $\Gamma_0$. Although the temperature is varied by more than two orders of magnitude, the slope only changes by a factor $\sim 3$. Within this limited range the decrease fits best with a logarithmic function of $\Gamma_\D/\Gamma_0$.

For almost all our data we have $\Gamma_\sub < \Gamma_\D < \Gamma_0$. The characteristic time corresponding to the Doppler width, $\Gamma_D^{-1}$, is shown in dashed lines in Fig.\ref{figexp_data}(b) for the lowest and highest temperatures. Since $\Gamma_\sub < \Gamma_\D$ it is not surprising that thermal motion affects subradiance, which is built up by the interference of light scattered by many atoms. However, it is a nonintuitive result that subradiant modes with lifetimes long compared to the typical atomic motion survive. A possible explanation for this apparent robustness is the large number of subradiant modes. As the atoms move and the eigenmodes of the system are modified, the excitation contained in a subradiant mode has a larger probability to stay in the subradiant manifold because, at large $b_0$, there are much more subradiant modes than superradiant ones \footnote{H. Ritsch, Workshop ``strongly disordered optical systems: from the white paint to cold atoms'', Carg\`ese (France), Sep. 2016}. In the extreme case of the Dicke limit ($R\leq \lambda$), there is only one superradiant mode in the low excitation limit, while $N-1$ subradiant modes are present.

\section{Temperature effects in the Coupled-Dipole model}
To provide a comparable numerical study of the decay dynamics of the interacting atomic ensemble, as well as to explore the regimes of parameters that we cannot reach experimentally, i.e. $\Gamma_\sub > \Gamma_\D$ and $\Gamma_\D > \Gamma_0$, we now turn to numerical simulations of an adapted version of the coupled-dipole (CD) model, which includes atomic motion  \cite{Bienaime2012,Eloy2018}.

The CD model provides a well-suited description in the context of cooperative effects, especially for super- and subradiance in the linear optics regime\,\cite{Javanainen1999,Svidzinsky2010,Courteille2010,Bienaime2011,Bienaime2012,Guerin2016a,Araujo2016,Roof2016,Araujo2018}.
The model consists of $N$ two-level atoms randomly distributed in space at position $\bm{r_i}$. The atoms are driven with an incident laser field with a Rabi frequency $\Omega(\bm{r_i})$ and a detuning $\delta$.
For the low excitation limit, which is considered here, the only relevant states are the ground state $|G\rangle = |g \cdots g \rangle$ and the single-excited states $|i\rangle = |g \cdots e_i \cdots g\rangle$.
One can then obtain an effective Hamiltonian for the time evolution of the atomic wave function
\begin{equation}
| \psi(t) \rangle = \alpha(t) | G \rangle +  \sum\limits_{i=1}^N \beta_{i}(t)| i \rangle \; . \label{eqpsi}
\end{equation}
Since we are in the low excitation limit ($\alpha \simeq 1$) the time evolution of the system is described by the time evolution of the excitation probabilities
\begin{equation}
\dot{\beta}_i = \left( i\delta-\frac{\Gamma}{2} \right)\beta_i -\frac{i\Omega_i}{2} + \frac{i\Gamma}{2} \sum_{i \neq j} V_{ij}\beta_j \; .
\label{eqbetas}
\end{equation}
The first term represents the single-atom decay dynamics, the second term the driving field for the excitation, and the last term the dipole-dipole interaction containing the cooperative effects.
We use a scalar model to describe the dipole-dipole interaction, neglecting any polarization effect as well as near-field terms, which is a good approximation for a very dilute gas, as used in our experiment.
In this case the dipole-dipole interaction term is
\begin{equation}
V_{ij} = \frac{e^{ik r_{ij}}}{k r_{ij}} \, , \mathrm{with} \; r_{ij} = |\bm{r}_i - \bm{r}_j| \; , \label{eqVij}
\end{equation}
with $r_{ij}$ the relative distances between the atoms.
Since the $\beta_i(t)$ provide the time evolution of the excitation probability one can then calculate the emitted light intensity as a function of time. A more detailed description can be found in Refs.~\cite{Bienaime2011,Araujo2018}.\\

\begin{figure}
\centerline{\includegraphics{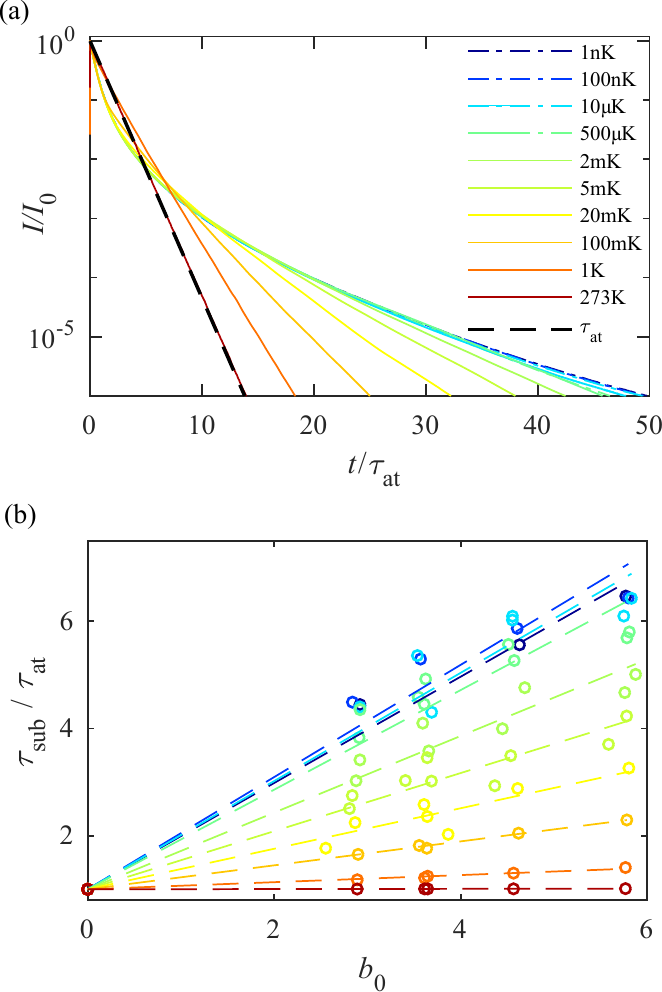}}
\caption{(a) Numerical results of the decay curves for $N_\text{at}=1500$ atoms. Subradiance starts to be significantly reduced in the mK range, but still shows decay time much longer than the single atom decay time $\tau_\text{at}$. Only when reaching room temperature does the decay curve follow the one of the single-atom case (dashed black line). The dashed lines are obtained with a ballistic motion of the atoms, while the solid lines are with the harmonic oscillator model. For \SI{10}{\micro \kelvin} and \SI{500}{\micro \kelvin} both models are displayed but are hardly distinguishable. (b) Numerical results for the fitted subradiant decay times for different resonant optical depths $b_0$. The temperatures are the same as in (a). For temperatures below \SI{2}{\milli\kelvin} the decay times are almost the same, while increasing the temperature further leads to a smooth decrease of the decay times. The dashed lines are linear fits, whose slopes $\alpha_\sub$ are reported in Fig.\,\ref{figdecay_t}.}
\label{figdecay_theo}
\end{figure}

In order to include the effect of temperature, we include atomic motion by assigning to each atom a velocity $v_i$ following a normal distribution of rms width $\sigma_v=(\kB T/\mRb)^{1/2}$ in each direction of space and let the space vector $\bm{r}_i(t)$ be time dependent.
At low temperature we use a ballistic motion 
for each atom, as in the experiment. However this turned out to be problematic with increasing temperatures, since it comes along with a non-negligible increase of size (and correspondingly a drop of the optical depth) of the sample during the subradiant decay time \,\footnote{This problem affects the results presented in the Fig.\, 3 of Ref.\,\cite{Bienaime2012}}. This effect is negligible in the experiment, where the spatial width of the distribution $\sigma_x\gg k\sigma_v\tau_\text{sub}$, but can become important in simulations, where one needs to use smaller values of $\sigma_x$ to simulate large values of $b_0$, as the number of atoms in the simulations is limited to a few 1000.
To avoid this and keep a well-defined optical depth during the simulation, for temperatures larger than \SI{500}{\micro \kelvin}, we simulate the atomic motion with a harmonic trapping. By choosing the oscillation frequency $\omega_\mathrm{H}$ as the ratio between the rms widths of the velocity and space distributions, $\omega_\mathrm{H}=\sigma_v/\sigma_x$, the sample keeps its Gaussian density distribution with a constant size. The initial position and velocity of each atom are drawn independently in their respective normal distributions.
We have checked that the two ways of including atomic motion (ballistic or harmonic) give the same results around \SI{500}{\micro \kelvin}, the temperature beyond which we use the harmonic motion.

\section{Scaling-law of the temperature dependence}
Solving Eqs.(\ref{eqbetas}) for different temperatures ranging from \SI{1}{\nano \kelvin} to \SI{273}{\kelvin} with an atom number $N=1500$ and computing the total scattered intensity, we obtain the decay curves shown in Fig.\,\ref{figdecay_theo}(a).
The long-lived subradiant states decay faster when the temperature increases; however, it reaches the single-atom decay only close to room temperature ($\Gamma_\text{D}/\Gamma_0\approx100$).
As in the experiment, we extract the subradiant decay time $\tau_\sub$ by an exponential fit of the slow decay at late time and low level (we choose the fitting range between $10^{-6}$ and $5 \times 10^{-5}$ for the relative intensity). The results are reported in Fig.\,\ref{figdecay_theo}(b) as a function of the on-resonance optical depth $b_0$. The linear trend of $\tau_\sub$ as a function of $b_0$ is well visible at all temperatures, which is consistent with the experimental observations.

Note that, at large temperature, the Doppler broadening becomes comparable to, or larger than, the detuning (we used $\delta = -4 \Gamma$ in all simulations), such that the driving laser is \textit{de facto} on resonance, which may introduce radiation trapping\,\cite{Labeyrie2003}. However, as studied in detail in Ref.\,\cite{Weiss2018}, radiation trapping dominates the slow decay only when the actual optical depth $b(\delta)$ is larger than $\sim 10$ at zero temperature, and is even less visible with Doppler-induced frequency redistribution. In our simulations, the Doppler-broadened optical depth is at most $\sim 0.5$, such that the observed slow decays in Fig.\,\ref{figdecay_theo} can be safely attributed to subradiance.

\begin{figure}
\centerline{\includegraphics{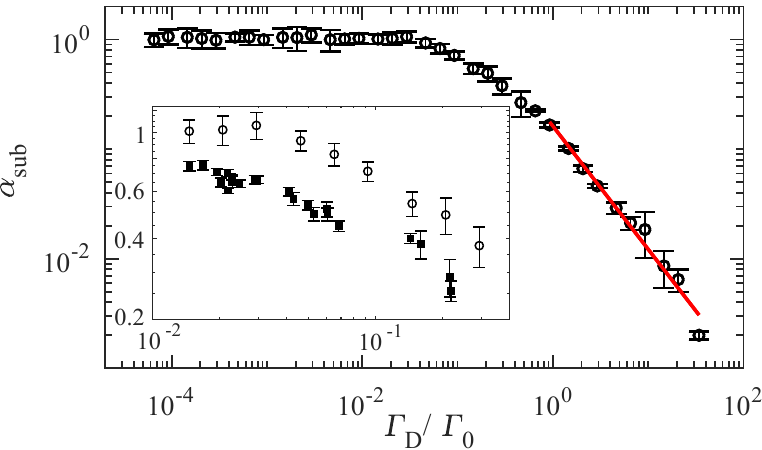}}
\caption{Numerical results for the subradiant decay time slopes $\alpha_\text{sub}$ as a function of the Doppler width $\Gamma_\D$ in units of $\Gamma_0$. After a plateau the decay time is reduced smoothly with increasing temperatures. When $\Gamma_\D/\Gamma_0>1$, the decay is a power law (note the log-log scales).  The fit (red solid line) gives an exponent compatible with -1. Inset: Direct comparison between the experimental (filled rectangles) and numerical (open circles) results for $\alpha_\sub$. The behaviors with the temperature are the same, which demonstrates the validity of our modeling of the atomic motion. }
\label{figdecay_t}
\end{figure}

To provide a more quantitative study on the impact of temperature on the subradiant decay times, we show in Fig.\,\ref{figdecay_t} the slopes $\alpha_\text{sub}$ extracted from the linear fit of the data of Fig.\,\ref{figdecay_theo}(b), as a function of the Doppler width in units of the natural linewidth. 
For the lowest temperatures a plateau is visible, i.e. $\alpha_\sub$ becomes independent of $T$ for low enough temperature, the atoms are quasistatic. When the Doppler broadening becomes nonnegligible, $\alpha_\sub$ starts to slowly decrease. This is the range we experimentally explore, as seen in the insets of Figs.\,\ref{figexp_data}(b) and \ref{figdecay_t}.
When the Doppler broadening reaches the single atom decay rate, $\Gamma_\D/\Gamma_0>1$, the decrease of subradiance follows a power law.
A power-law fit $\alpha_\text{sub}=\beta (\Gamma_\D/\Gamma_0)^m$ gives $\beta \simeq 0.16$ and $m=-1.12\pm 0.14$, compatible with $m=-1$. 
This exponent can be interpreted as follows. In the regime when $\Gamma_\D/\Gamma_0>1$, the convolution of the scattering cross-section with the Doppler broadening leads to a reduced center of line scattering cross section scaling as $(\Gamma_\D/\Gamma_0)^{-1}$ and, for a given atomic density and sample size, the optical depth is also proportional to $(\Gamma_\D/\Gamma_0)^{-1}$.
The observed scaling of the subradiant decay rates as shown in Fig.\,\ref{figdecay_t} is thus primarily consistent with a Doppler-broadened resonant optical depth in the limit of $\Gamma_\D \gg \Gamma_0$ .
Note that a simple selection of slow atoms up to a certain cutoff velocity inside the Maxwell-Boltzmann distribution would lead to a different scaling, with a more drastic decrease as $(\Gamma_\D/\Gamma_0)^{-3}$. 

We finally show in the inset of Fig.\,\ref{figdecay_t} the direct comparison (without any free parameter) of the measured and computed slopes $\alpha_\sub$ as a function of the temperature. Although we do not expect any quantitative agreement on the precise values of $\alpha_\sub$, even at zero temperature, because of the complex level structure of rubidium that is not taken into account in the model, the behaviors of $\alpha_\sub$ with $T$ are in remarkable agreement. This validates \textit{a posteriori} our way of introducing the atomic motion in the coupled-dipole model.

\section{Conclusion and Outlook} \label{secsummary}
In summary, we have demonstrated that for a large temperature range of the atomic cloud the subradiant decay is robust against thermal decoherence. In particular, the time scale corresponding to the Doppler broadening, $\Gamma_\D^{-1}$, does not directly introduce a limit for the subradiant lifetime, but merely provides a rescaling of the subradiant enhancement factor ($\alpha_\text{sub} b_0$) in the high temperature regime. This rescaling can be interpreted as a modification of the resonant optical depth, which is reduced by the Doppler broadening but remains the cooperativity parameter controlling subradiance.

These results open up the prospect of observing and using subradiance at room temperature or with hot atomic vapors. Indeed, an extrapolation of the scaling laws discovered in this work predicts subradiance with $\tau_\sub/\tau_\at \sim 40$ with a \SI{5}{\centi\meter} Rb cell at \SI{100}{\celsius} (with these parameters, $b_0 \sim 10^4$ and $\Gamma_\D/\Gamma_0 \sim 40$ \cite{Baudouin2014a}). Similar to room-temperature atomic quantum memories, which have already reached excellent performances~\cite{Katz2018,Guo2019}, subradiance could be more broadly used for quantum-optics or quantum-metrology applications~\cite{Ostermann2013,Scully2015,Facchinetti2016,Asenjo2017,Facchinetti2018,Guimond2019} at room temperature. This robustness may also be applied to other interference effects in light scattering, such as coherent back scattering \cite{Cherroret2019}.

\begin{acknowledgments}
\section*{Acknowledgments}
The authors thank Romain Bachelard for useful discussions and advice on the numerical simulations.
This work was supported by the ANR (project LOVE, Grant No. ANR-14-CE26-0032), the European Training Network ColOpt (European Union Horizon 2020 program under the Marie Skodowska-Curie action, Grant agreement No. 721465), the Brazilian Coordena\c{c}\~ao de Aperfei\c{c}oamento de Pessoal de N\'{i}vel Superior (CAPES), and the Deutsche Forschungsgemeinschaft (Grant No. WE 6356/1-1).
\end{acknowledgments}


%

\end{document}